\documentclass[12pt]{iopart}

\usepackage{iopams}
\usepackage{graphicx}
\usepackage{dcolumn}
\usepackage{bm}

\begin{document}

\title{Stripe structures in phase separated magnetic oxides}

\author{O.~G.~Udalov$^{1,2}$, I.~S.~Beloborodov$^{1}$}
\address{$^1$ Department of Physics and Astronomy, California State University Northridge, Northridge, CA 91330, USA}
\address{$^2$ Institute for Physics of Microstructures RAS, Nizhny Novgorod, 603950, Russia}
\ead{oleg.udalov@csun.edu}

\vspace{10pt}

\begin{indented}
\item[]April 2019
\end{indented}

\begin{abstract}
We investigate the phase separated inhomogeneous charge and spin states in magnetic oxides.  In particular, we study one dimensional harmonic waves and stripe structures. We show that harmonic spin charge  waves are unstable and inevitably transform into two or three dimensional structures, while the stripe structures can be stable for certain parameters. Such stripe structures may allow the control of magnetic state with electric field in a magnetic oxide thin film.
\end{abstract}

%
\vspace{2pc}

\section{Introduction}\label{Sec:Intro}

Magnetic oxides (MO) are materials attracting attention of numerous scientific groups due to their intriguing physics and strong many-body effects. Magnetic properties of these materials are defined by several phenomena such as super-exchange, electronic correlations, Jahn-Teller effects, orbital and charge ordering and etc. Numerous homogeneous phases are well known in magnetic oxides~\cite{Dunaevskii2004, PhysRevLett.85.3954,PhysRevLett.82.1016}. Among them are ferromagnetic (FM), antiferromagnetic (AFM) A, G, CE phases, canted states, etc.

At that, enormous attention were paid to studying the non-uniform states in magnetic oxides both theoretically and experimentally~\cite{DAGOTTO20011,DAGOTTO20039,1367-2630-7-1-067,doi:10.1002/pssb.2221860102,1063-7869-38-5-R02}. The inhomogeneous states are considered to be the key feature of these materials and are closely related to explanation of the colossal magneto-resistance effect in these materials.

The inhomogeneous states appear in the form of regular charge-ordered (CO) structures ~\cite{PhysRevLett.83.5118,PhysRevB.62.9432,PhysRevLett.83.5118} as well as
in the form of the random intermixture of two different phases (FM conductive and AFM insulating)~\cite{PhysRevB.80.024423,PhysRevLett.64.475,PhysRevB.53.8671}

Regular charge-ordered (stripe or checkerboard) structures with a period of few (two) lattice parameters appears at high (close to half filling) electron doping due to the combination of Coulomb repulsion, Jahn-Teller effects and orbital effects. Wider stripes   occurs at lower concentration \cite{Cheong1998,Littlewood2005}.  In this case the stripe structure has larger period (up to 5 lattice parameters). Appearance of the stripe structure in this case is attributed to the Jahn-Teller effect.

Inhomogeneous states in MOs are also predicted even in the absence of the Coulomb repulsion and Jahn-Teller effect in the double exchange model with classical and quantum spins~\cite{Kagan1999, doi:10.1002/pssb.2221860102,PhysRevB.78.155113}. This is so-called phase separation, meaning that canted AFM regions shrink into FM regions with higher electron concentration surrounded by the AFM insulating areas. Since the FM regions with increased electron concentration are charged, the long-range Coulomb interaction is important for formation of the charge separated states. In  \cite{Nagaev1990,Nagaev1974,1063-7869-38-5-R02} the Coulomb interaction was taken into account by the separation of the crystal into spherical Wigner cells. The influence of the Coulomb potential on the electron wave functions was not taken into account.

The ultimate phase separated state is the polaron state in which each electron forms small (a few lattice constants) FM cluster around it ~\cite{1063-7869-44-6-R01,0305-4470-36-35-304,Dagotto257,doi:10.1002/cphc.200600188,Moreo2034,Nagaev1968}. Such a state exists at low electron concentration until polaron percolation occurs.

Phase separated states with broad range of characteristic scale from few nm to microns were reported in various experimental works~\cite{PhysRevB.63.172419}.

Phase separation is important for the colossal magneto-resistance effect in magnetic oxides. Ordinarily it is assumed that the regions with different phases are randomly distributed across the sample~\cite{PhysRevLett.93.037203}.
At low concentration the FM regions do not overlap forming a network of independent metallic regions separated by an insulating matrix. Conductivity enhances greatly in the vicinity of percolation of FM conductive regions. In this case the infinite metallic cluster may appear through the whole sample. In the vicinity of percolation the system becomes extremely sensitive to the external magnetic field.

Recently, MOs attract much of attention as magneto-electric (ME) materials. There are intrinsic magneto-electric effects due to spin-orbit interaction~\cite{PhysRevLett.98.057601} as well as spin-charge-orbital coupling~\cite{vandenBrink2008}. ME coupling was also studied due to strain and charge accumulation in hybrid systems MO/ferroelectric~\cite{PhysRevB.87.094416, PhysRevB.75.054408}.
The important question exists if ME coupling can occur due to the phase separation.

The inhomogeneous magnetic states in MOs are often related to the charge density inhomogeneities. This opens the way to control magnetic structure of the materials with electric field. One can imagine that charged FM regions can be moved across the sample under the action of external electric field. Recently, local control of phase separated states with electric field was discussed in ~\cite{doi:10.1080/00150193.2017.1292823}. The enhancement of magneto-electric effect
due to phase separation was shown in \cite{Hernandez2015}. This
adds another merit to investigate inhomogeneous states in MOs.

Random 3-dimensional (3D) intermixture of charged FM regions in AFM matrix is not the best object to control with external electric field. The more appropriate object for interaction with electric field is the 1-dimensional (1D) stripe charge-spin structures. However, the stability of stripe structures is questionable. The 3D inhomogeneous structures should be more energetically favorable taking into account the long-range Coulomb repulsion. In the present manuscript we will study 1D inhomogeneous structures and their stability in MO taking into account the long-range Coulomb interaction. We demonstrate that these structures can be stable under certain conditions.

The manuscript is organized as follows. In section~\ref{Sec:model} we introduce the model. Next we briefly describe the homogeneous phase in MO within the proposed model. In section~\ref{InhomogeneousAnalytics} we consider our model analytically and study the stability of 1D inhomogeneous phases. Section~\ref{Sec:Numerical} is devoted to numerical investigation of stripe structures in MOs. Finally, in section~\ref{Discussion} we discuss possible implications of 1D stripe structure and propose how one can control magnetic state with electric field in a phase separated MO.

\section{Main results}

Here we provide our main results which will be discussed in the rest of the manuscript.

1) At low electron concentration ($<5\%$) the polaron state is the most favorable.

2) At higher electron concentration $>5\%$ the macroscopic inhomogeneous states may appear
in MO depending on the system parameters instead of a uniform state.

3) Harmonic 1D charge-spin waves are more favorable than the uniform state in a certain parameter range.
However, these states are not stable against the 2D and 3D perturbations and cannot survive in the system.

4) The 1D stripe structure is more favorable than the uniform state.
At that these states are stable against the 2D and 3D perturbations. We suggest that these
states can be used to realize the magneto-electric effect in MOs.

\section{The model}\label{Sec:model}

The system Hamiltonian has the form
\begin{equation}\label{Eq:Ham}
\hat H=-\sum_{<i,j>} t_{ij}\hat a^+_i \hat a_j+\mathrm{C.C.}+J\sum_{<i,j>}\bi{S}_i\bi{S}_j+\hat H_{\mathrm{C}},
\end{equation}
where $\bi{S}_i$ is the ``classical'' magnetic moment (normalized) of $i$-site, $J>0$  is the (AFM) intersite exchange coupling, $\hat a_i$ and $\hat a^+_i$ are the creation and annihilation operators for an electron at the site $i$, $t_{ij}$ is the transfer matrix element. This element depends on the mutual orientation of magnetic moments of sites $i$ and $j$, $t_{ij}=t\mathrm{cos}(\theta_{ij}/2)$, where $\theta_{ij}=\widehat{\bi{S}_i\bi{S}_j}$. Note that summation in both terms is performed over the nearest neighbours. We assume here cubic lattice. Therefore, each site has 6 neighbours. The last term describes the  Coulomb interaction.

There are several types of magnetic states in the system: 1) uniform state with $\theta_{ij}$ independent of coordinates; 2) polaron state, in which electrons are strongly localized and form a small 3D perturbation of magnetic structure. Within the polaron region the ions magnetic moments form canted FM state with $\theta_{ij}<\pi$. All the space outside the polarons is in the AFM state. 3) Charge and spin waves, where electron and spin density periodically change along a certain direction in the system. In the rest of the work we will consider these types of structures.

\subsection{Simplified model}

Let us consider the case of extremely low electron concentration and large scale density waves. In this case we can assume that locally electrons are in the media with constant $\theta_{ij}=\theta$. Moreover, all electrons are at the bottom of the conduction band with the kinetic energy

\begin{equation}\label{Eq:ElEnBottomBand}
E_\mathrm{el}=-6t\mathrm{cos}(\theta/2).
\end{equation}
Local energy (the energy per one cell) without the Coulomb interaction is given by

\begin{equation}\label{Eq:LocEn}
E_\mathrm{loc}=-6t\mathrm{cos}(\theta/2)n+6J\cos^2(\theta/2),
\end{equation}
where $n$ is the unitless number of electrons per one cell. The first term describes the kinetic energy of the electrons and the second term is the magnetic energy. Note that we count the energy from the energy of AFM state.

The normalized average magnetization per one unit cell in the system is exactly $m=\cos(\theta/2)$. So, we can write
\begin{equation}\label{Eq:LocEnM}
E_\mathrm{loc}=-6tmn+6Jm^2.
\end{equation}

\section{Uniform state}\label{Sec:Unif}

In the homogeneous state we can neglect the Coulomb interaction since the charge density is zero (negative electron charges are
compensated by positive ion charges). The magnetization $m$ is uniform across the system. According to (\ref{Eq:LocEnM}) in this case we have either uniform FM state or canted FM (or canted AFM) state. Magnetization $m_\mathrm{un}$ and total energy $E^\mathrm{tot}_\mathrm{un}$ are given by
\begin{equation}\label{Eq:Unif}
m_\mathrm{un}=\left\{\eqalign{\frac{tn_0}{2J},~\frac{tn_0}{2J}<1,\cr 1,~\frac{tn_0}{2J}>1,}\right.~
E^\mathrm{tot}_\mathrm{un}=\left\{\eqalign{-\frac{3}{2}\frac{t^2n_0^2}{J},\frac{tn_0}{2J}<1, \cr 6J-6tn_0,~\frac{tn_0}{2J}>1.}\right.
\end{equation}
Here parameter $n_0$ is the value of uniform electron concentration. Further we will use this notation to denote the
average electron concentration or $N/N_\mathrm{sites}$ ($N$ is the total number of electrons
in the system and $N_\mathrm{sites}$ is the total number of sites in the system). One can see that for low electron density (or large $J$,
or small $t$) the system is in the canted FM (or canted AFM) state. At high electron density the system switches to FM state. Generally, conducting electrons push the system to the FM state, while the exchange interaction between the ions favors the AFM state.

Below we use the uniform state as a reference one. We will calculate the
energy gain due to various inhomogeneous states with respect to corresponding energy of the uniform state.

\section{Inhomogeneous states}\label{InhomogeneousAnalytics}

The logic of the following consideration is the following. First, we start with low electron concentration.
In this case the polaron state is the most energetically favorable. However, these states cannot be used
to realize the magneto-electric coupling. Therefore, next we will consider higher electron concentrations
where macroscopic inhomogeneities can occur. We consider 1D type of structures.
As a first example, we treat the spin-charge harmonic waves. These states as we will show are not stable against 2D
perturbations and can not survive. Therefore, we will study another type of structures, namely the
1D stripe structures which are stable.
In this section we discuss a simplified analytical model of these inhomogeneous states and
in section~\ref{Sec:Numerical} we will perform the numerical modeling of stripe structures in MOs.

\subsection{Polaron state}

First, we consider the case of very low electron concentration, $n\ll 1$. In this case the system may form an inhomogeneous state consisting of single magnetic polarons. Consider a system without electrons. The AFM intersite interaction leads to the
formation of checkerboard AFM state. Adding an electron into some site leads to
rotation of magnetic moment of this site. On one hand such a rotation allows electron hopping to neighbouring sites and therefore it decreases the electron kinetic energy. On the other hand rotation of magnetic moment increases the magnetic energy. The competition of electron kinetic and ion magnetic energies defines the system ground state. We introduce the angle between the magnetic moment of the site $i$ and its
neighbouring sites as $\theta\ne \pi$. Angles between all other magnetic moments are $\pi$. Thus, electron can not hop beyond the nearest neighbours of the site $i$ and we have a polaron consisting of 7 sites. The Hamiltonian describing this system is given by
\begin{equation}\label{Eq:HamRed}
\hat H_{\mathrm{reduced}}=\left(\begin{array}{ccccccc}
0 -&\tilde t &-\tilde t& -\tilde t &-\tilde t &-\tilde t &-\tilde t \\
-\tilde t &0 &0 &0 &0 &0 &0 \\
-\tilde t& 0& 0& 0& 0& 0& 0 \\
-\tilde t& 0& 0& 0& 0& 0& 0 \\
-\tilde t& 0& 0& 0& 0& 0& 0 \\
-\tilde t& 0& 0& 0& 0& 0& 0 \\
-\tilde t& 0& 0& 0& 0& 0& 0
\end{array}\right)-6J\mathrm{cos}(\theta),
\end{equation}
where $\tilde t=t\mathrm{cos}(\theta/2)$. The lowest energy state (polaron state) is $E_\mathrm p=-\sqrt{6}t\mathrm{cos}(\theta/2)+6J\mathrm{cos}(\theta)$. Minimizing the energy with respect to $\theta$
one finds that the polaron energy and angle are given by
\begin{equation}\label{Eq:PolStates}
\eqalign{
&\mathrm{cos}(\theta_\mathrm p/2)=\frac{t}{4\sqrt{6}J},~E_\mathrm p=-\frac{t^2}{8J}, t<4\sqrt{6}J,\\
&\mathrm{cos}(\theta_\mathrm p/2)=1,~E_\mathrm p=-\sqrt{6}t+12J, t>4\sqrt{6}J.}
\end{equation}
As before we calculate the energy with respect to the AFM state energy.
The first line corresponds to the canted FM state of the polaron. The bottom line is for FM ordering.
The total energy density of polaronic state is the single
polaron energy multiplied by the electron density $E_\mathrm p^\mathrm{tot}=nE_\mathrm p$.

Note, that increasing $t$ (decreasing $J$) should lead to formation
of bigger polarons. The energy of bigger polarons can be found in a similar away. One can start with the smallest polaron of 7 sites ((0,0,0), ($\pm$1,0,0), (0,$\pm$1,0) and (0,0,$\pm$1) sites) and add the site which is in the corner of the initial polaron ((1,1,0) site, for example). Then one gets bigger polaron consisting of 12 sites.
Next we can add another sites and get the polaron with sizes of 16, 20 24 sites etc. We compare the energy of polarons with different size depending on the exchange coupling $J$ and found that the small polaron is the most favorable if $J>0.33t$. Decreasing $J$ below this critical value leads to fast increase of the polaron size. ($J<0.033t$ - 12 sites, $J<0.027t$ - 16 sites, $J<0.017$ - 24 sites, $J<0.009$ - 28 sites etc.).

\subsubsection{The Coulomb interaction of polarons}

Here we estimate the gain in the Coulomb energy due to formation of the polaron state. Consider a macroscopic system with $N\gg 1$ electrons and volume $\Omega$. The Coulomb interaction operator is given by $\hat H_\mathrm C=\sum_{i\ne j}(1/|\bi{r}_i-\bi{r}_j|)$, where $i$ and $j$ enumerate electrons. Consider at first the system with delocalized electrons with the wave function uniformly distributed across the whole volume. Such a system corresponds to the uniform magnetic state considered previously. The average Coulomb energy in this case is given by $E_\mathrm C^\mathrm{un}\approx (e^2/(2\varepsilon ))(N^2-N)(1/\Omega^2 )\int\int_{\Omega}drdr'(1/(|\bi{r}-\bi{r}'|))$. The last integral is of order of $1/R$, where $R$ is the system linear size. $-N$ correction is due to subtraction of the electron self-interaction energy. Lets now estimate the Coulomb energy of the polaronic state. In this case we have $E_\mathrm C^\mathrm{pol}\approx (e^2/(2\varepsilon))\sum\sum_{i\ne j}(1/|\bi{r}_i-\bi{r}
_j|)$. To estimate the sum we transform it into an integral considering that far from a certain electron the average electron concentration is $N/\Omega$, but within the distance $a$ from an electron there are no other electrons. Here $a$ is the average spacing between the polarons. Finally we get $E_\mathrm C^\mathrm{pol}\approx (e^2/(2\varepsilon))(N^2-N)(1/\Omega^2 )(\int\int_{\Omega}drdr'(1/(|\bi{r}-\bi{ r}'|))-\int\int_{|\bi{r}-\bi{r}'|<a}drdr'(1/(|\bi{r}-\bi{r}'|)))$. The last term
can be considered as a correction to the main term. This correction is of order of $e^2N^2a^2/R^2\sim (e^2/a)N\sim e^2N^{4/3}$ (here
we use the fact that $\Omega/a^3=N$). In the case of many electrons this correction is much bigger than that due to the self-interaction in the case of homogeneous system. Therefore, the polaronic state has lower Coulomb energy than the homogeneous state. The energy gain per one site can be estimated as
\begin{equation}\label{Eq:CoulEnGainPol}
\Delta E_\mathrm C^\mathrm{pol}\approx-2\pi ne^2/(\varepsilon a).
\end{equation}
This equation has a clear physical meaning since
electrons in the polaronic state are located at distance $a$ from each other. So, each electron is surrounded by positive charge within the radius $a$. The energy of interaction of the electron with this positive charge is of order of $2\pi e^2/(\varepsilon a)$.

\subsubsection{Percolation problem}

Single polaron includes 7 sites. Therefore, for concentration of carriers more than $n>1/7\approx 14$\% the system is fully covered by polarons and one can not consider the system as an ensemble of independent polarons. In fact the percolation appears much earlier when covered volume is of order of 1/3. We can estimate the percolation threshold on the level of $n\approx 5$\%. For low exchange coupling, $J<0.03t$ the polaron size grows fast leading to decreasing the percolation threshold. As discussed in~\cite{Altshuler2006} the delocalization can also occurs due to the electron-electron scattering. This could reduce the percolation threshold even more.

Finally, we conclude that below the  percolation threshold the polaronic state is the most favorable and formation of one dimensional spin-change waves or other large scale inhomogeneous structures (with positive contribution from the Coulomb term) are unlikely.
Above the threshold concentration the model of delocalized electrons is more appropriate. We consider this model in the next section.

\subsection{Harmonic charge density waves}

Let us now assume that electron concentration is high enough and polarons overlap forming delocalized electron wave functions.  To describe the system we will follow the simplified model in (\ref{Eq:LocEnM}). Consider 1D harmonic spatial variation of electron density and magnetic moment. Assume that the angle $\theta$ (and therefore the magnetization) harmonically varies in space.
\begin{equation}\label{Eq:Mvar}
m=m_0+m_1\cos(kx).
\end{equation}
The electron density oscillates as well
\begin{equation}\label{Eq:Nvar}
n=n_0+n_1\cos(kx),
\end{equation}
producing the charge density
\begin{equation}\label{Eq:Nvar1}
\rho=\rho_1\cos(kx)=e(n-n_0)/\delta^3.
\end{equation}
The lattice constant is $\delta$ giving the volume of the unit cell $\delta^3$.
In the case of harmonic oscillations the magnitude $n_1$ can not exceed the doping level $n_0$.
The charge variations produce the Coulomb contribution to the
total system energy (averaged over the period and per one site)
\begin{equation}\label{Eq:EnHamWave1}
E^{\mathrm{tot}}_\mathrm{1D}=6Jm_0^2-6tm_0n_0+3Jm_1^2-3tm_1n_1+\frac{U_0 n_1^2}{(k\delta)^2},
\end{equation}
where $U_0=\pi e^2/(\varepsilon\delta)$ is the characteristic Coulomb interaction of two electrons sitting at neighbouring sites.

We can exclude parameter $t$ from (\ref{Eq:EnHamWave1}) considering energy being normalized by $t$.
We introduce $\tilde J=J/t$ and $\tilde U_0=U_0/t$. For simplicity, we omit the sign $\sim$
in further consideration assuming that we measure energy $E$, $J$ and $U_0$ in $t$.
Then (\ref{Eq:EnHamWave1}) takes the form
\begin{equation}\label{Eq:EnCDW2}
E^{\mathrm{tot}}_\mathrm{1D}=6Jm_0^2-6m_0n_0+3Jm_1^2-3m_1n_1+\frac{U_0 n_1^2}{(k\delta)^2},
\end{equation}

Similarly, we consider 3D fluctuations with
\begin{equation}\label{Eq:Mvar3D}
\eqalign{
&m=m_0+m_1\cos(kx)\cos(ky)\cos(kz),\\
&n=n_0+n_1\cos(kx)\cos(ky)\cos(kz).}
\end{equation}
The energy in this case is given by
\begin{equation}\label{Eq:EnCDW3D}
E^{\mathrm{tot}}_\mathrm{3D}=6Jm_0^2-6m_0n_0+\frac{3}{4}Jm_1^2-\frac{3}{4}m_1n_1+\frac{U_0 n_1^2}{12(k\delta)^2}.
\end{equation}

\subsubsection{Stability against 1D and 3D perturbations}

Here we study the stability of the uniform state considered
in section~\ref{Sec:Unif} against 1D and 3D spin-charge waves.
We substitute $m_0=m_\mathrm{un}$ into equations~(\ref{Eq:EnCDW2}) and (\ref{Eq:EnCDW3D}). Consider first the uniform canted state.
It is stable ($E^\mathrm{tot}_\mathrm{1D}(m_1,n_1)$ and has a minimum at $m_1=0$ and $n_1=0$)
against the appearance of the 1D wave when the following criterion is satisfied
\begin{equation}\label{Eq:UnifCrit1D}
\frac{U_0J}{(k\delta)^2}>\frac{3}{4}.
\end{equation}
Similarly one gets for the criterion of stability against the 3D perturbation
\begin{equation}\label{Eq:UnifCrit3D}
\frac{U_0J}{(k\delta)^2}>\frac{9}{4}.
\end{equation}

Comparing equations~(\ref{Eq:UnifCrit1D}) and (\ref{Eq:UnifCrit3D}) one can see that the region of instability of the uniform state against the 3D fluctuation is bigger than the region for 1D perturbations. This means that if such a system starts with a uniform state,
the 3D structure (not 1D) will always develop. However, this does not mean that if one
prepares the inhomogeneous 1D state, this state will inevitably decay into 3D structure. 1D state can be a
metastable (as for example, single ferroelectric domain state is metastable but can be transformed
into the multidomain ground state by application of an electric field).  This is related to the fact that considered system is non linear. In the next sections we consider two types of 1D structures
and study their stability against developing of 2D inhomogeneity.

Note that for the FM state ($m_\mathrm{un}=1$) the perturbation should be slightly modified $m_0=m_\mathrm{un}-m_1$ since $m$ should be less than 1. In this case the linear in $m_1$ terms appear in the energy with positive coefficient, meaning that
uniform FM state is stable against weak perturbations.
\begin{figure}
\includegraphics[width=0.5\columnwidth]{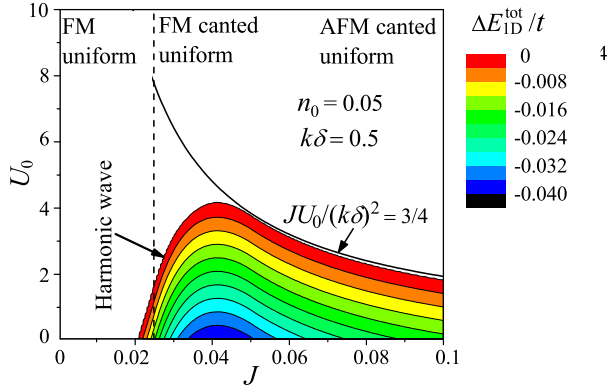}
\caption{Energy gain $\Delta E^\mathrm{tot}_{1\mathrm D}/t$ (normalized by the characteristic kinetic energy $t$) due to harmonic 1D spin-charge wave with $m_0=\mathrm{min}(m_1,1-m_1)$. White color means that the uniform state is more favorable. $J$ is the exchange interaction (normalized by $t$) and $U_0$ is the characteristic Coulomb interaction energy (normalized by $t$).}   \label{Fig:PDw2}%
\end{figure}

\subsubsection{Harmonic 1D wave with high amplitude.}

There are no terms in the energy that restrict growth of variation amplitudes $m_1$ and $n_1$. Therefore, if the uniform state is unstable then the amplitudes $m_1$ and $n_1$ increase until some of them reaches a maximum possible value.  Therefore, here we consider strong 1D fluctuations of magnetic moment with $m_1=m_0$ if $m_0<1/2$ (or $m_1=1-m_0$ if $m_0>1/2$). We minimize the energy in
(\ref{Eq:EnCDW2}) with respect to $m_1$ and $n_1$  in the region $n_1<n_0$ and $m_0<1/2$. After that we find the energy gain, $\Delta E^\mathrm{tot}_\mathrm{1D}=\mathrm{min}(E^\mathrm{tot}_\mathrm{1D})-E^\mathrm{tot}_\mathrm{un}$.

Figure~\ref{Fig:PDw2} shows the energy gain as a function of parameters $U_0$ and $J$. The region where the state with 1D spin-charge wave is more favorable than the uniform state is shown with colors. White color shows the region where the uniform state is the ground state. Solid line shows the 1D stability criteria in (\ref{Eq:UnifCrit1D}). Dashed line shows the separation of uniform FM state and canted uniform state. Note, that in a small part of the FM uniform region the 1D spin-charge wave is more favorable. So, while the FM uniform state is stable against the small perturbation (as mentioned in the previous section), there are high amplitude perturbation having lower energies. One needs to overcome some energy barrier to get into these inhomogeneous states.

Let us now consider if the 1D spin-charge wave is stable against the 2D perturbations. We consider the magnetization and charge variations of the following form
\begin{equation}\label{Eq:Var2D}
\eqalign{
&m_\mathrm{2D}=m_1(1+\cos(kx)(1+\delta_\mathrm m\cos(ky)),\\
&n_\mathrm{2D}=n_0+n_1\cos(kx)(1+\delta_\mathrm n\cos(ky)).}
\end{equation}
Following the same procedure as we used previously one can get that strong 1D fluctuations are not stable against 2D perturbation in the region $JU_0<3t^2/2$. This means that in the whole region where strong 1D wave is more favorable than the uniform state the 1D perturbation will transform into 2D or 3D inhomogeneous structure.

Note that the 2D perturbation in the form of (\ref{Eq:Var2D}) is valid only for
$m_1=m_0\ne 0.5$. For $m_1=m_0=0.5$ a slightly different perturbation should be taken.
However, the 1D state is still unstable in this case.

\subsection{Stripe perturbation}

In this section we consider another type of perturbations of the uniform state - stripe structure. The stripe structure has a period $L=l\delta$. We assume that all electrons are concentrated in the part of the period $0<x<d\delta$. Below we will measure the distance in interatomic spacings, $\delta$. In the rest of the period there are no electrons and magnetic state is AFM. Electron concentration and magnetic moment in the stripe structure are given by
\begin{equation}\label{Eq:Meandr}
\left\{\eqalign{&n_\mathrm s=\frac{n_0 l}{d},~ m_\mathrm s=\frac{n_0tl}{2Jd},~~ 0<x<d,\\ &n=0,~ m=0,~~ d<x<l.}\right.
\end{equation}

\begin{figure}
\includegraphics[width=0.5\columnwidth]{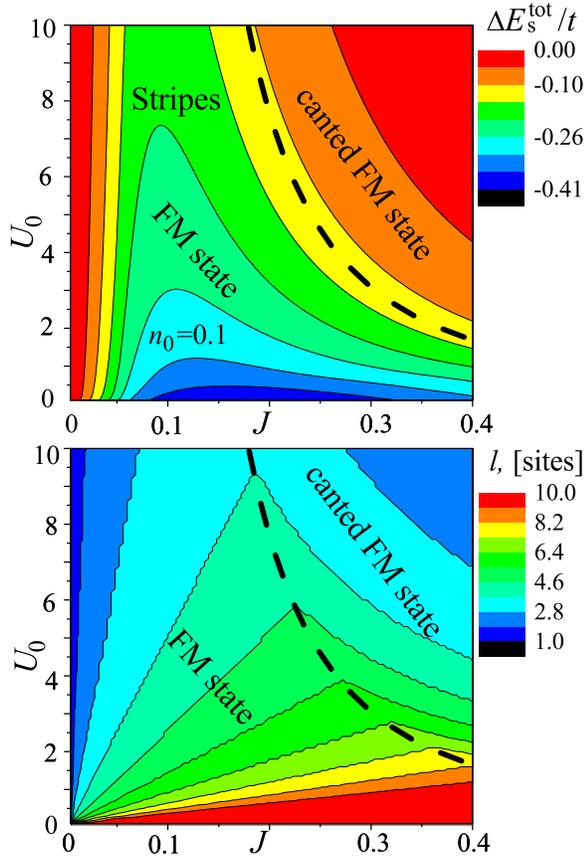}
\caption{Upper panel: energy gain due to the stripe perturbation $\Delta E^\mathrm{tot}_{\mathrm s}/t$ (normalized by $t$). Lower panel: the stripe period at which the maximum gain is achieved. Plots correspond to the initial electron concentration $n_0=0.1$. Black dashed line shows the boundary in the parameter space between the states with FM ordering ($m_\mathrm s=1$) in the electron enriched region and canted FM ordering ($m_\mathrm s<1$).  $J$ is the exchange interaction (normalized by $t$) and $U_0$ is the characteristic Coulomb interaction energy (normalized by $t$).  \label{Fig:Stripe1}}%
\end{figure}

First we consider if the stripe structure is more energetically favorable than the uniform state and
find parameters region where this happens. The total system energy is given by
\begin{equation}\label{Eq:EnStripe1}
E^{\mathrm{tot}}_\mathrm s=6Jm^2d/l-6tmn_0+l^2U_0n_0^2(1-d/l)^2/6.
\end{equation}
One can see that the first two terms in (\ref{Eq:EnStripe1}) depend on $l/d$. The Coulomb energy is positive and is a function of $l/d$ multiplied by $l^2$. If one keeps the ratio $l/d$ constant but decreases the period of the stripe structure $l$, then the first two terms stay the same. At that the Coulomb interaction decreases as $l^2$ meaning
that the smaller the period the smaller the system energy.
In this model nothing restricts the system from decreasing the stripe period. The local electron energy depends only on the electron density and not on the stripe period, while the Coulomb interaction decreases with reducing the period of charge oscillations. However,
the region where all electrons seat cannot be smaller than the single atomic plane. So, we have a restriction on $d$, $d>1$. There is also upper bound for $l$ at a given $d$. Since the electron concentration should be less than 1, then $l<d/n_0$. Finally, we should work in the region $d>1$, $l/d>1$, $l/d<1/n_0$. Due to the discussed properties of the energy $E^{\mathrm{tot}}_\mathrm s$, the minimum appears at $d=1$ and $l>1$, $l<1/n_0$.

Note that in real systems electron energy increases with decreasing parameter $d$.
This factor is not taken into account in this consideration. There are also other factors that produce lower bound for $d$. In section~\ref{Sec:Numerical} all these factors will be taken into account. In sections~\ref{Sec:Kdep} and ~\ref{Sec:kin} we also discuss how the simplified model can be extended to avoid shrinking of the stripe structure.

Here we assume that electrons are all in a single atomic plane and $d=1$. Then the only parameter we have is the stripe period $l$. Minimizing the system energy over $l$ one can find the energy gain due to the stripe structure, stripe period, electron concentration and magnetization in the electron enriched area.

Figure~\ref{Fig:Stripe1} shows the energy gain due to the stripe structure comparing to the uniform state as a function of parameters $J$ and $U_0$. The parameter region corresponds to real materials constants: hopping matrix element $t=0.1 - 0.4$ eV, the intersite Coulomb interaction $U_0 = 1-10$ eV, and $J=0.01 - 0.1 t$.

One can see that in the whole range of $J$ and $U_0$ the stripe structure is more favorable than the uniform state. The energy gain reaches $0.4t$ for low $U_0$. Lower panel shows the optimized period of the stripe structure corresponding the maximum energy gain. One can see that at low $U_0$ the period reaches its possible maximum $l=10$ for the given electron concentration $n_0=0.1$. Increasing $U_0$ leads to the reduction of
the stripe structure period. Black dashed line shows the boundary between regions where $m_\mathrm s=1$ and $m_\mathrm s<1$. For small $J$ and $U_0$ the region $0<x<d$ is in FM state, for large $U_0$ and $J$ there is a canted FM state in this area. The line is described by the equation
\begin{equation}\label{Eq:FMboundary}
U_0=\frac{9n_0}{2J(2J-n_0)}.
\end{equation}

Let us now consider a question if the stripe structure is
stable against 2D perturbations. First we consider the following type of perturbation
\begin{equation}\label{Eq:Var2DStripe}
\eqalign{
&m_\mathrm{2D}=\left\{\eqalign{&m_\mathrm s(1-\tilde m_1+\delta_\mathrm m\cos(ky)),~0<x<d,\\ &0,~ d<x<l,}\right.\\
&n_\mathrm{2D}=\left\{\eqalign{&n_\mathrm s(1+\delta_\mathrm n\cos(ky)),~0<x<d,\\&0,~ d<x<l.}\right.}
\end{equation}
where
\begin{equation}\label{Eq:Var2DStripe1}
\eqalign{
&\tilde m_1=0,~ m_\mathrm s<1,\\
&\tilde m_1=\delta_\mathrm m,~ m_\mathrm s=1.}
\end{equation}
This is a wave of magnetization and charge in the FM region $0<x<d$. The cases of FM ordering $m_\mathrm s=1$ and canted FM state should be treated separately. This is related to the fact that $m<1$ and therefore to get the wave in the FM region one needs to decrease the average magnetization.

We calculated the Coulomb contribution to the energy due to such a perturbation (see details in the Appendix). It is given by
\begin{equation}\label{Eq:Coulomb2DStripe1}
E_\mathrm C=\frac{U_0n_0^2(l-d)^2}{6}+\frac{U_0n_\mathrm s^2\delta_\mathrm n^2d}{2(k\delta)^2l}.
\end{equation}
The first term here is the energy of the unperturbed stripe structure and the second term is the correction due to 2D perturbation, (\ref{Eq:Var2DStripe}). This correction is positive and  proportional to the perturbation amplitude $\delta_\mathrm m$ squared.

First consider the case with $m_\mathrm s<1$ and assume that we are away
from the boundary (\ref{Eq:FMboundary}). In this case the energy corrections due to the perturbation are the following $+3Jm_\mathrm s^2\delta_\mathrm m^2-3m_\mathrm sn_0\delta_\mathrm m\delta_\mathrm n+n_\mathrm s^2\delta_\mathrm n^2dU_0/(2(k\delta)^2l)$. The system is stable against the 2D perturbation when
$JU_0>3(k\delta)^2/2$. Increasing $k$ decreases the stability of the stripe structure in the region $m_\mathrm s<1$. Since $k\delta<2\pi$ then the stripe structure would be stable for $JU_0>6\pi^2$. This, however, realizes well beyond the parameter region we studied.
Therefore, we can conclude that the system with canted FM state is not stable against the 2D perturbation.

Let us now consider the case with $m_\mathrm s=1$ (parameter region below the line (\ref{Eq:FMboundary})).
In this case the energy correction is different. An
additional term appears, $12Jd/l\delta_\mathrm m(n_0l/(2Jd)-1)+3J\delta_\mathrm m^2-3n_0\delta_\mathrm m\delta_\mathrm n+n_\mathrm s^2\delta_\mathrm n^2d/(2(k\delta)^2l)$
(here $m_\mathrm s=1$). The most important term here is the first one, linear in $\delta_\mathrm m$.
Moreover, the coefficient in front of $\delta_\mathrm m$ is positive. The energy linearly grows
due to 2D perturbation. Nonlinear in $\delta_\mathrm m$ and $\delta_\mathrm n$ terms can be neglected while the perturbation is small.
This means that the 1D stripe structure with $m_\mathrm s=1$ is stable against the 2D perturbation considered above. It is possible that for
strong enough 2D perturbation (where non linear terms are more important than linear one) the 2D structure can develop.
However, one should overcome an energy barrier to destroy the 1D structure and create 2D one. This behavior is the consequence of non linear behavior of the magnetization as a function of concentration. Thus, below the line (\ref{Eq:FMboundary}) the 1D stripes are stable against perturbations (\ref{Eq:Var2DStripe}).

If one comes close to the boundary  (\ref{Eq:FMboundary}) from the
FM side then the energy barrier protecting the 1D state reduces. We will have mostly 1D structure with weak 2D variations. Going deeper into the region beyond the boundary  (\ref{Eq:FMboundary}) the 2D variations increases and eventually the system would be totally 2D or 3D inhomogeneous.

Let us now consider another type of 2D perturbation. In particular, periodically bended stripe described by
\begin{equation}\label{Eq:Var2DStripeBend}
\eqalign{
&m_\mathrm{2D}=\left\{\eqalign{&m_\mathrm s,~\delta_\mathrm b \cos(ky)<x<d+\delta_\mathrm b \cos(ky),\\ &0,~ d+\delta_\mathrm b \cos(ky)<x<l+\delta_\mathrm b \cos(ky),}\right.\\
&n_\mathrm{2D}=\left\{\eqalign{&n_\mathrm s,~0+\delta_\mathrm b \cos(ky)<x<d+\delta_\mathrm b \cos(ky),\\&0,~ d+\delta_\mathrm b \cos(ky)<x<l+\delta_\mathrm b \cos(ky).}\right.}
\end{equation}
One can show that the local energy $E_\mathrm{loc}$ does not change under these variations since the volume where electrons gather together does change. A correction appears to the Coulomb interaction energy due to the bending $E_\mathrm C\approx U_0n_0^2(l-d)^2/6+U_0n_\mathrm s^2\delta_\mathrm b^2d/l$ (see calculations in the Appendix). One can see that the correction to the Coulomb energy is positive. Therefore, this kind of perturbation is not energetically favorable, meaning that 1D stripe structure is stable against them. Finally, we can conclude that the 1D stripe structure can be stable if the magnetization in the electron enriched region is saturated.

\subsection{Extension of the simplified model. Adding gradient terms}\label{Sec:Kdep}

As we discussed in the previous section the stripe structure favours the smallest possible $d$
and there is no factor restricting this in the model (besides artificially introduced limit of a single interatomic distance). To overcome this difficulty one can introduce spatial derivatives into the model
\begin{equation}\label{Eq:LocEnGrad}
E_\mathrm{loc}=-6mn+6Jm^2+\delta^2\kappa_\mathrm m(\nabla m)^2+\delta^2\kappa_\mathrm n(\nabla n)^2.
\end{equation}
From phenomenological point of view it is natural to introduce such terms into an inhomogeneous system. The gradient term is positive and prevents formation of waves in the system. First, these gradient terms will smear the boundary between FM and AFM states in the stripe structure.
In our previous consideration there was an abrupt interface between these two phases. Second, these gradient terms will
prevent decreasing of the stripe period. Decreasing the stripe period inevitably increases the energy associated with the gradient terms.
We do not consider this extension since in the next section we study the stripe structures using numerical simulations where
all ``gradient terms'' are included automatically.

\subsection{Extension of the simplified model. Taking electron spectrum into account}\label{Sec:kin}

Previously, we assume that all electrons are at the bottom of the conduction band and have the same energy $E_\mathrm{el}=-6t$. One can consider a more general model with parabolic electron spectrum $E_\mathrm{el}=-6t+t(k_x^2+k_y^2+k_z^2)$. Taking such a spectrum into account one makes formation of the inhomogeneous states less favorable. If one increases the electron density in a certain region of space the kinetic energy of electrons in this region will be larger than in the model considered in the previous sections.

Due to the parabolic electron spectrum the local energy in (\ref{Eq:LocEn}) acquires an additional positive term $+9.1tmn^{5/3}$. Generally, this term can be treated analytically, while increasing the complexity of all equations. One can show that the system becomes more stable against all possible perturbations due to this correction. For example, 1D stripe structures become more stable against the 2D perturbations.
At that this correction does not protect harmonic 1D waves from 2D perturbations.

What, however, important is that the additional term depends only on the ratio $d/l$ in the case of stripe structure. Therefore, it does not help with the problem of unrestricted decreasing of $d$ in the simplified model.

In the next section we treat the initial Hamiltonian in
(\ref{Eq:Ham}) numerically and show how the phase diagram of stripe structures changes comparing to the simplified model.

\section{Numerical modeling}\label{Sec:Numerical}

As we discussed in the previous section the simplified model has two limitations: i) the absence of factor limiting shrinking of stripe period;
and ii) over simplified  electron band structure which takes into account
only the lowest energy level. At high electron concentration (more than 5\%) this leads to underestimate
of electron kinetic energy. In this section we perform numerical modeling taking into account
the kinetic energy in a correct way. Also our modeling contains factors preventing decreasing of the stripe period.
So, one can optimize the stripe period in contrast to the simplified model considered in the previous section.

\subsection{Modeling procedure}

Here we study the system described by the Hamiltonian in (\ref{Eq:Ham}). Let us introduce the notations $\bi{r}=(x,y,z)$.
Coordinates $x$, $y$, and $z$ are measured in units of lattice spacing. The electron transfer is possible between neighbouring
sites only. We introduce the notation $t^{\pm x,y,z}_\bi{r}$ standing for the matrix element of the transfer from the position $\bi{r}=(x,y,z)$ to the neighbouring site along the axis shown in the superscript $^x$, $^y$ or $^z$ in the positive $^+$ or negative $^-$ direction. There are corresponding angles $\vartheta^{\pm x,y,z}_\bi{r}$ between magnetic moments of relevant neighbouring sites.

We consider a periodic system with a period of $l$ sites in the x-direction and $\vartheta_{(x+l,y,z)}^{\pm x,y,z}=\vartheta_{(x,y,z)}^{\pm x,y,z}$. The system is uniform in the (y,z) plane and $\vartheta_{\bi{r}}^{\pm x,y,z}$ does not depend on $y$ and $z$.

Wave functions of electrons are considered as plane waves with a wave vector $\bi{k}$ ($\Psi_\bi{k}=(\psi^{1}_\bi{k},~\psi^{2}_\bi{k},~...,\psi^{l}_\bi{k})^Te^{i(k_x x+k_y y+k_z z)}$). $\psi^{x}_\bi{k}$ is the wave function amplitude at the site $i$. Introducing these solutions into the Hamiltonian in (\ref{Eq:Ham}) one can get the $l$ by $l$ matrix equation for the amplitudes $\psi^x_\bi{k}$. In our calculations the $k$-space is divided into 100$\times$100$\times$100 parts.
\begin{figure}
	\includegraphics[width=0.5\columnwidth]{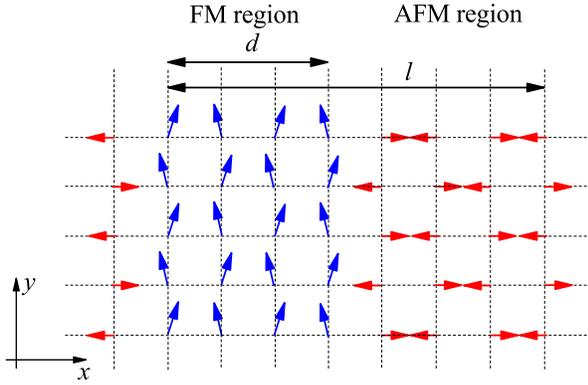}
	\caption{Stripe perturbation. Blue and red arrows show magnetic moments at sites. $l$ is the period of the stripe structure, $d$ is the width of canted FM region enriched with electrons. \label{Fig:Meandr}}%
\end{figure}

The Coulomb interaction is taken into account via a self-consistent procedure. This procedure was used previously for modeling of MO/FE and MO/I interfaces and MO's superlattices~\cite{PhysRevB.84.155117, PhysRevB.73.041104,PhysRevB.80.125115,PhysRevB.78.024415}.
We fix the electron concentration $n_0$ (number of electrons $N$) during our calculations. At each step we calculate the electron density  distribution $\rho_\bi{r}$ and then the distribution of electric potential, $\Phi_\bi{r}$. Both the density and the potential are independent of $y$ and $z$. Therefore, the 1D discrete version of the Maxwell equation is $\Phi_{x+1}-2\Phi_x+\Phi_{x-1}=-4\pi\delta^2(\rho_x-\rho_0)/\varepsilon$, where $\rho_x=e\sum_\bi{k} |\psi^{x}_\bi{k}|^2$, $\rho_0=|e|n_0$ is the positive charge density (assumed to be uniform). Summation is over all occupied states. We consider the case of zero temperatures. So, only $N$ lowest levels are occupied. The obtained electrical potential is used in the next step, for calculating the electron wave functions. These steps are performed until the potential and electron density converge to a certain solution.
Finally, we find the total energy consisting of electron energy and magnetic energy.

\subsection{Uniform system}

First we study the uniform system describing by a single parameter, $\vartheta^{\pm x,y,z}_{\bi{r}}=\vartheta_0$. We minimize the total system energy over $\vartheta_0$ and find the ground state. Note, that the Coulomb interaction does not play any role for
uniform system, since the electron density is uniform and there is no excessive charge in the system and no electric field.
\begin{figure*}[t]
	\begin{center}
		\includegraphics[width=1.0\columnwidth]{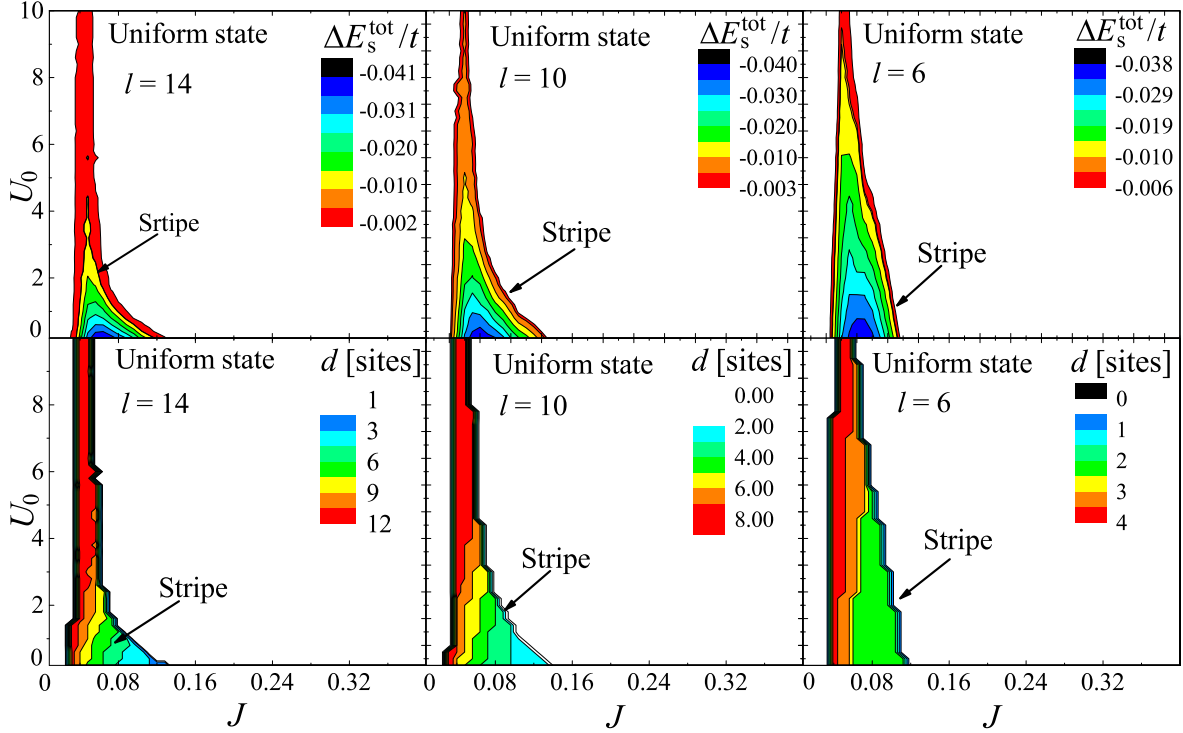}
		\caption{Numerical calculations. Upper row: the normalized energy gain $\Delta E^\mathrm{tot}_\mathrm s/t$ (normalized by $t$) due to the stripe structure for given electron concentration $n_0=0.1$ and different stripe period $l=14,~10,~6$ interatomic distances. $U_0$ is the characteristic Coulomb interaction energy (normalized by $t$) and $J$ is the exchange interaction between localized moments (normalized by $t$). Lower row: the length of the electron enriched region $d$ at which maximum energy gain is reached. \label{Fig:NumStripe1}}%
	\end{center}
\end{figure*}

\subsection{Polaron states and harmonic charge waves}

As we mentioned before the polaron states cannot be used to realize the magneto-electric coupling. Therefore, we do not perform numerical simulations of such states here. Also we skip modeling of 1D harmonic spin-charge
waves since they are unstable against 2D perturbations as we showed in the previous section.

\subsection{Stripe structure}

The stripe structure (shown in figure~\ref{Fig:Meandr}) is defined by the period $l$ and the width of the electron enriched region $d$.
For $0\le x\le d-1$ the angle between the magnetic moments is less than $\pi$ (canted or FM state). For electron concentration $n_0<d/l$ all electrons are located in this region. In the region $d+1\le x\le l-2$ the AFM ordering is realized. There are two interface layers ($x=d$ and $x=l-1$) connecting regions with different magnetic state. The stripe structure is described by the following spatial distribution of angles
\begin{equation}\label{Eq:Stripe}
\eqalign{
&\theta_\bi{r}^{+x}=\left\{\eqalign{&\mathrm{acos}\frac{n_0l}{2Jd},~\frac{n_0l}{2Jd}<1,\\&0,~\mathrm{overwise}}\right\},~0\le x\le d-1,\\
&\theta_\bi{r}^{+x}=\frac{\pi}{2}+\frac{1}{2}\left\{\eqalign{&\mathrm{acos}\frac{n_0l}{2Jd},~\frac{n_0l}{2Jd}<1,\\&0,~\mathrm{overwise}}\right\},~x=d,l-1,\\
&\theta_\bi{r}^{+x}=\pi, d+1\le x\le l-2,\\
&\theta_\bi{r}^{\pm y,z}=\left\{\eqalign{&\mathrm{acos}\frac{n_0l}{2Jd},~\frac{n_0l}{2Jd}<1,\\&0,~\mathrm{overwise}}\right\},~0\le x\le d-1,\\
&\theta_\bi{r}^{\pm y,z}=\pi, d\le x \le l-1.}
\end{equation}

\begin{figure*}[t]
	\begin{center}
	\includegraphics[width=1.0\columnwidth]{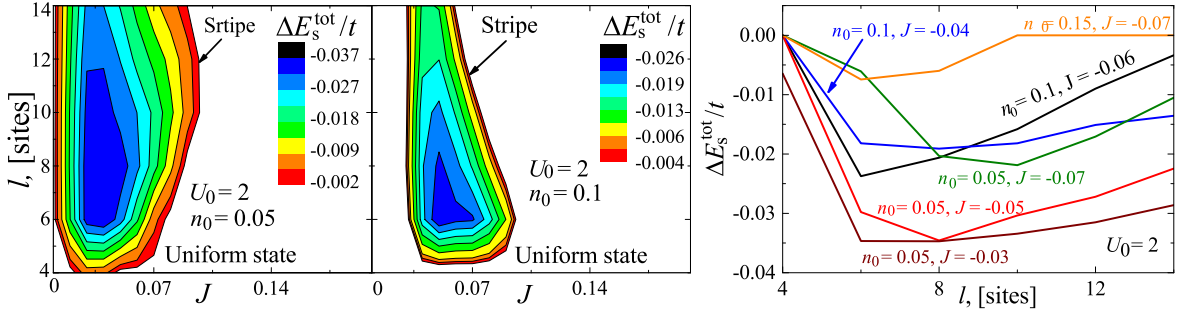}
		\caption{Numerical calculations. Left and central panels: energy gain $\Delta E^\mathrm{tot}_\mathrm s/t$ (normalized by $t$)
		due to a stripe structure as a function of stripe period $l$ and exchange constant $J$ at given Coulomb interaction $U_0=2$ and electron concentrations $n_0=0.05$ (left panel) and $n_0=0.1$ (central panel). Right panel: energy gain due to the stripe structure as a function of $l$ at $U_0=2$ and different combinations of $J$ and $n_0$ parameters. \label{Fig:NumStripeLJ}}%
	\end{center}
\end{figure*}

In contrast to analytical treatment in the previous section, here we do not have any arguments allowing us to chose $d$. Therefore, we follow
a different approach. We first calculate energy gain due to the stripe structure at a given stripe period $l$ and various $d$. We find the maximum energy gain varying $d$ for fixed $l$. After that we perform similar calculation for different $l$ and define the most energetically favorable structure at a given $J$, $U_0$ and $n_0$.

Figure~\ref{Fig:NumStripe1} shows the region in the parameter space where the stripe structure is energetically more favorable. Upper panels show the energy gain due to the stripe perturbation comparing to the uniform state for electron concentration $n_0=0.1$ and different stripe period $l=14,~10,~6$. Lower panels demonstrate the electron enriched region size $d$ at which maximum energy gain occurs. White color indicates that the uniform state is more energetically favorable. One can see that the parameters region in which the stripe structure can be realized is much smaller comparing to what was obtained in the simplified analytical model. This is mostly due to the fact that electrons are not at the bottom of the band (as was assumed in the simplified model). This becomes especially important in the electrons enriched region. Therefore, the electron kinetic energy is underestimated in the simplified model. One can also see that energy gain due to stripe structure is smaller than what was obtained in the simplified model. One can see from the bottom panel that electron enriched region shrinks as we decrease the Coulomb interaction $U_0$. Electron enriched area is a half of the period (and even less) at low $U_0$.

Next figure shows how the energy gain depends on $l$ and $J$ at a given $U_0$ and $n_0$. The left and central panel demonstrate two dimensional diagrams for $U_0=2$ and $n_0=0.05$ and $n_0=0.1$, correspondingly. White color shows the parameters region where the uniform state is more favorable. One can see that the region where the stripe structure may exist becomes smaller as we increase the electron concentration. For example, for $U_0=2$ the stripe structures are not favorable at all when electron concentration is more than 0.2.

Figure~\ref{Fig:NumStripeLJ} shows that there is a non monotonic dependence of the energy gain on $l$.
The gain decreases as $l$ becomes large or small enough. Therefore, there is certain optimal stripe structure period. Right panel shows dependence of the energy gain due to stripe structure as a function of single parameter $l$ for different combinations of $U_0$, $J$, and $n_0$. One can see that all curves have a minimum energy corresponding to some finite optimal period $l$. The optimal period decreases with increasing of the electron concentration. At $n_0=0.5$ the optimal period reaches 10 iteratomic spacings (see green line), at $n_0=0.1$ the period is 8 sites (see blue line) and at $n=0.15$ the period decreases to 6 sites. The optimal period depends on $J$ as well. For example, it grows from 6 sites at $J=-0.03$ (brawn line) to 10 sites at $J=-0.07$ (green line). Our calculations show that the optimal period decreases with increasing $U_0$.

\section{Discussion}\label{Discussion}

\begin{figure}
\includegraphics[width=0.5\columnwidth]{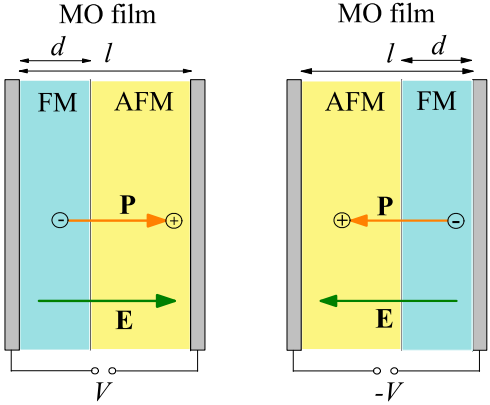}
\caption{Magnetic oxide thin film inside a capacitor. 1D stripe structure is realized in the MO under applied external electric field. Depending on the direction of the electric field $\bi{E}$ the FM regions stay at the left or at the right surface of the film. Electric polarization $\bi{P}$ appears in the MO film. \label{Fig:MEeffect}}
\end{figure}

1) In the introduction section we mention that one can control the 1D stripe structure with electric field.
Figure~\ref{Fig:MEeffect} shows the idea. Consider the MO film with the thickness corresponding to the optimized period of 1D stripe structure. Applying an electric field to this film induces such a 1D structure with the FM charged region located at the corresponding film surface. Applying the opposite electric field one can switch position of the FM and AFM regions and move the FM region to the other side. Thus, one can control the magnetic state of the film with electric field. Important question here is what happens when we switch off the electric field. According to our findings the 1D structure can be stable in the system. However, the FM region can move into the middle of the film or can
stay at the film edge. In the first case there is no electrical polarization in the film. In the second case a non-zero remnant electrical polarization occurs. FM region position depends on the boundary conditions at the film interface. This question requires a separate investigation.

2) Our model does not take into account the Jahn-Teller effect. It  leads to electron localization. Adding the Janh-Teller effect into the model should increase the stability of 1D stripe structure.

3) Considered model does not take into account different orbital states at the same site. Introducing several orbitals on the same site would make the system even more complicated. For example, instead of two (FM and AFM) phases there can appear other phases coexisting in the same system.

4) We do not take into account disorder in MOs. This disorder can appear due to random positions of dopands. This randomness can destroy the regular structure studied in this work. This issue requires an additional investigation.

\section{Conclusion}

We studied inhomogeneous charge and spin states in MOs within the one band double-exchange model.
We treated this model analytically neglecting electronic band structure.
Also we used numerical simulations to study inhomogeneous states in MOs.
We mostly concentrated on 1D structures such as harmonic spin-charge waves and stripe structures.
We showed that 1D harmonic waves are not stable against 2D and 3D perturbations and
can not survive in MOs (at least in our model). At that, the stripe structure is stable
against 2D perturbations and can exist in the system. Using numerical simulations we defined
parameters region where these structures may exist and defined the optimal period of the stripes.
We showed that such stripe structures can be used to realize the magneto-electric effect in MO thin films.

\ack

This research was supported by NSF under Cooperative Agreement Award EEC-1160504.
O.~U. was supported by the Foundation for the
Advancement of Theoretical Physics and Mathematics BASIS (grant 18-1-3-32-1) and Russian Foundation for Basic Researches (Grant  18-32-20036).

\appendix
\section*{Appendix. The Coulomb interaction in the perturbed stripe structure}
\setcounter{section}{1} 

Here we calculate the Coulomb energy of stripe structures perturbed by a 2D wave. First, consider the perturbation
in (\ref{Eq:Var2DStripe}). In this case the potential in two regions (region (1): $d<x<l$, region (2): $0<x<d$)  can be written as follows
\begin{equation}\label{Eq:potential}
\eqalign{
\Phi^{(1)}=&\phi^{(1)}_1(x-d)-2U_0n_0(x-d)^2+\\&+\phi^{(1)}_{3+}\cos(ky)e^{-k(x-d)}+\phi^{(1)}_{3-}\cos(ky)e^{k(x-d)},\\
\Phi^{(2)}=&\phi^{(2)}_1(x-d)+2U_0n_0\frac{l-d}{d}(x-d)^2+\\&+\phi^{(2)}_{3+}\cos(ky)e^{-k(x-d)}+\phi^{(2)}_{3-}\cos(ky)e^{k(x-d)}.
}
\end{equation}
Using the boundary conditions $\Phi^{(1)}=\Phi^{(2)}|_{x=d}$, $\Phi^{(1)}|_{x=l}=\Phi^{(2)}|_{x=0}$, $\partial \Phi^{(1)}/\partial x=\partial \Phi^{(2)}/\partial x|_{x=d}$, and $\partial \Phi^{(1)}/\partial x|_{x=l} =\partial \Phi^{(2)}/\partial x|_{x=0}$ one can find all coefficients in (\ref{Eq:potential}). Then the Coulomb interaction per unit of volume is calculated as $E_\mathrm C= k/(4l\pi)\int dx\int dy \Phi(\bi{r})\rho(\bi{r})$, where $\rho$ is the charge density defined in regions (1) and (2) as follows $\rho^{(1)}=n_0/\delta^3$ and $\rho^{(2)}=-(n_\mathrm s-n_0)/\delta^3$.

In the case of bending perturbation in (\ref{Eq:Var2DStripeBend}) the solution is more complicated.
We follow the approach proposed in~\cite{KUZNETSOV2019104}. We calculate the Coulomb energy assuming
that the bending is small $\delta_\mathrm b\ll d,l$ and $2\pi/k$. The electrical potential has the form
\begin{equation}\label{Eq:potential1}
\eqalign{
\Phi^{(1)}=&-2U_0n_0(x-d)(x-l)+\\&+\phi^{(1)}_{3+}\cos(ky)e^{-k(x-d)}+\phi^{(1)}_{3-}\cos(ky)e^{k(x-d)},\\&~d+\delta_\mathrm b\cos(ky)<x<l+\delta_\mathrm b\cos(ky)\\
\Phi^{(2)}=&2U_0n_0\frac{l-d}{d}(x-d)x+\\&+\phi^{(2)}_{3+}\cos(ky)e^{-k(x-d)}+\phi^{(2)}_{3-}\cos(ky)e^{k(x-d)},\\&~\delta_\mathrm b\cos(ky)<x<d+\delta_\mathrm b\cos(ky).
}
\end{equation}
To obtain quadratic in $\delta_\mathrm b$ corrections to the Coulomb energy one needs to find the coefficients $\phi^{(1,2)}_{3\pm}$ linear in $\delta_\mathrm b$. We use the boundary conditions $\Phi^{(1)}=\Phi^{(2)}|_{x=d+\delta_\mathrm b\cos(ky)}$, $\Phi^{(1)}|_{x=l+\delta_\mathrm b\cos(ky)}=\Phi^{(2)}|_{x=\delta_\mathrm b\cos(ky)}$, $\partial \Phi^{(1)}/\partial x=\partial \Phi^{(2)}/\partial x|_{x=d+\delta_\mathrm b\cos(ky)}$, and $\partial \Phi^{(1)}/\partial x|_{x=l+\delta_\mathrm b\cos(ky)} =\partial \Phi^{(2)}/\partial x|_{x=\delta_\mathrm b\cos(ky)}$ to find linear in $\delta_\mathrm b$ potential. Finally, we calculate the Coulomb energy in the same way as in the previous case.

\section*{References}
\bibliography{CDW}

\end{document}